\documentstyle[aps,preprint]{revtex}
\begin{document}
\draft

\title{The Scalar-Tensor Inflationary Cosmology}
\author{Mian WANG }

\vspace{3mm}

\address{Department of Physics, Henan Normal University.\\
          Xinxiang, Henan, China  453002}
%\date{today}
\maketitle
\def\be{\begin{equation}}
\def\ee{\end{equation}}
\def\sk{\vspace{3mm}}
\def\bsk{\vspace{1cm}}
\def\hs{\hspace{3mm}}
\begin{abstract}
In this paper the scalar-tensor theory of gravity is assumed to
describe the evolution of the universe and the gravitational
scalar $\phi$ is ascribed to play the role of inflaton. The theory
is characterized by the specified coupling function $\omega(\phi)$
and the cosmological function $\lambda(\phi)$. The function
$\lambda(\phi)$ is nearly constant for $0<\phi<0.1$ and
$\lambda(1)=0$. The functions $\lambda(\phi)$ and $\omega(\phi)$
provide a double-well potential for the motion of $\phi(t)$.
Inflation commences and ends naturally by the dynamics of the
scalar field. The energy density of matter increases steadily
during inflation. When the constant $\Gamma$ in the action is
determined by the present matter density, the temperature at the
end of inflation is of the order of $10^{14} GeV$ in no need of
reheating.  Furthermore, the gravitational scalar is just the cold
dark matter that men seek for.

\bsk
PACS:\ \ \ 04.20.-q;\ \ \ 98.80.Cq.

\end{abstract}

\newpage
\section{INTRODUCTION}

The idea of inflation $\cite{Guth}$, as well known, gives not only
the solution for the horizon problem and flatness problem that
plagued the standard big bang model, but also the explanation of
the formation of large-scale structure of the universe from the
evolution of the primordial density perturbations. According to
current approach, during the evolution of the very early universe,
one usually assumes that the energy density happens to be
dominated by some vacuum energy and then comoving scales grow
exponentially or quasi-exponentially, and the vacuum energy
driving inflation is generally assumed to be associated to some
scalar field, the inflaton, which is displaced from the minimum of
its potential. There results a dilemma. The level of density and
temperature fluctuations observed in the present universe,
$\delta\rho/\rho$ $\sim 10^{-5}$, require the inflaton potential
to be extremely flat, and this is in contrast with the requirement
that the coupling of the inflaton field cannot be too small
otherwise the reheating process, which converts the vacuum energy
into radiation at the end of inflation, takes place too slowly and
insufficiently.$\cite{Lidsey}$

\sk

The futile effort to identify the inflaton in particle sector
and the fact that gravitational force is
the only long-range force governing the evolution of the universe
%and dominating the formation of the large-scale structure of the universe
lead us to consider the inflation as a pure gravity effect and the
gravitational scalar in the scalar-tensor gravity of Bergmann,
Nordtvedt and Wagoner $\cite{Bergmann}$ as the inflaton field.
\sk

In section \ref{stgt}, we set down the basic formalism. In
the action of the scalar-tensor gravity theory, there are a
coupling function $\omega(\phi)$ and a cosmological function
$\lambda(\phi)$. The function $\lambda(\phi)$ is nearly constant
for $0<\phi<0.1$ and $\lambda(1)=0$. The functions $\lambda(\phi)$
and $\omega(\phi)$ provide a double-well potential for the motion
of $\phi(t)$. In the Lagrangian there are three constants $\xi$,
$\beta$ and $\Gamma$ in $\omega(\phi)$, $\lambda(\phi)$ and the
coupling term respectively. The $\xi$ and $\beta$ are determined
by the large-scale structure data, while $\Gamma$ is determined by
the matter content of the universe, i.e., the temperature of CMB
and the density of non-relativistic matter together. We shall have
a model with $(\xi, \beta, \Gamma)=(7.5,\, 1.7 \times 10^{-16},\,
0.06362)$. The basic cosmological equations for the Robertson-Walker
metric are given in section \ref{cosmo}.

\sk

In section \ref{infl} we discuss the
very early universe. The universe is created from nothing but gravity.
After a quantum era,
% about which we know very little,
we set the initial
condition at the Planck time. The inflation commences naturally
because of the pressure $p_\phi$ of the scalar field $\phi$ is negative
and equals in magnitude with the density $\rho_\phi$ of $\phi$.
The gravitational scalar field $\phi$ is just the inflaton field.

\sk

In section \ref{after}, we discuss the cosmological solutions after inflation.
Section \ref{model} proposes a realist model. Section \ref{nucleo} discusses
the primordial nucleosynthesis.

\bsk
\section{THE SCALAR-TENSOR GRAVITY THEORY}\label{stgt}
We have proposed such a scalar-tensor gravity theory
$\cite{Wang}$. It is characterized by an action:
\be
{\cal A}=\int d^4x \sqrt{-g}\,\,\left[-\phi R - \frac{\omega}{\phi} \phi\cdot\phi
-2 \phi \lambda(\phi) -\frac{\Gamma (u\cdot\phi)^2}{1-\phi}+16 \pi L_m\right] ,
\label{st1}
\ee
in which $L_m$ is the Lagrangian of matter,
$\phi\cdot\phi\equiv\phi_{,\sigma} \phi^{,\sigma}$,
 $u\cdot\phi\equiv u_{\mu}\phi^{,\mu}$,
 $u_{\mu}$ is the four-velocity, $\phi^{,\mu}\equiv
\partial\phi/\partial x_{\mu}$, and $\Gamma$ is a constant of
 mass dimension 0. The $\Gamma-$term describes the coupling of the
 field $\phi$ with matter which is described as an ideal fluid.
The coupling function $\omega(\phi)$ and the cosmological function
$\lambda(\phi)$ are given as
\be
2\omega(\phi)+3 = \frac{\xi}{1-\phi}\, ,
\label{st2}
\ee
\be
\lambda(\phi) = 2\xi \beta (1-\phi- \phi \,\, ln\,\phi) ,
\label{st3}
\ee
where $\xi $ and $\beta$ are two dimensionless constants.
In a model, we can set  $\beta=1.7\times 10^{-16}$ and
$\xi=7.5$ by the data of COBE DMR observations
(we shall see in section \ref{model}).
This value  of $\beta$ is extremely small indeed,
yet there is no fine-tuning problem, since it is not a parameter in
gauge theory but a number to characterize the cosmological function
$\lambda$.
We note that the function $\lambda(\phi)$ is nearly constant of
value $2\xi\beta$ for $\phi<0.1$ and tend to 0 at $\phi=1$.
That is to say, in the very early universe, there is a very large
cosmological constant $\Lambda \sim O(10^{-15})$, but at present
day, it will be very small since $\phi\,\rightarrow\,1$.
Here we adapt the natural units: $\hbar =c=G=1$.

\sk

Varying action $A$ with respect to $g^{\mu\nu}$, we have field
equations
\be
R_{\mu\nu}-\frac{1}{2}\,g_{\mu\nu}R=-\frac{8\pi}{\phi}\,T_{\mu\nu}
\label{st4}
\ee
where $T_{\mu\nu}$ is the total stress tensor $T_{\mu\nu}\equiv t_{\mu
\nu}+\tau_{\mu\nu}$. The $t_{\mu\nu}$ and $\tau_{\mu\nu}$ are the
stress tensors of matter and field $\phi$ respectively, they read
\be
t_{\mu\nu}=p_m g_{\mu\nu}+(\rho_m+p_m)\, u_\mu u_\nu
\label{st5}
\ee
and
\begin{eqnarray}
\tau_{\mu\nu}=\frac{\omega}{8\pi\phi}\left(\phi_{,\mu}\phi_{,\nu}-\frac{
 g_{\mu\nu}}{2}\phi\cdot\phi\right)&+&\frac{1}{8\pi}\left(\phi_{;\mu\nu}
 -g_{\mu\nu}(\Box\phi+\phi\lambda)\right) \nonumber \\
 &+& \frac{\Gamma}{8\pi(1-\phi)}
 \left(2u_\mu\phi_\nu u\cdot\phi-\frac{g_{\mu\nu}}{2}(u\cdot\phi)^2\right) ,
\label{st6}
\end{eqnarray}
where $\rho_m$ and $p_m$ are energy density and pressure of matter
respectively.
Varying the action $A$ with respect to $\phi$, we have field equation
of $\phi$:
\be
(2\omega+3)\Box\phi=8\pi t^{\mu}_{\,\,\mu}+2\phi^2\lambda'-2\phi\lambda
-\omega'\phi\cdot\phi-\frac{\Gamma\dot\phi^2}{(1-\phi)^2}-\frac{2\Gamma
\phi\ddot\phi}{1-\phi}\, .
\label{st7}
\ee

\bsk

\section{THE COSMOLOGY: BASIC EQUATIONS}\label{cosmo}

According to the principle of cosmology, $\phi=\phi(t)$.
Thus in the Robertson-Walker metric
\be
ds^2=-dt^2+a^2(t)\left(\frac{dr^2}{1-kr^2}+r^2d\theta^2+r^2sin^2\theta
d\phi^2\right) ,
\label{frw}
\ee
there results the following equations:
\be
H^2\equiv
\left(\frac{\dot {a}}{a}\right)^2 =
 \frac{\lambda}{3}+\frac{\omega \dot\phi^2}
 {6\phi^2}-\frac{\dot a}{a}\frac{\dot\phi}{\phi}
 +\frac{8\pi}{3\phi}\rho_{m}-\frac{\Gamma}{2\phi}\frac{\dot\phi^2}
{1-\phi}-\frac{k}{a^2}\, ,
\label{ab1}
\ee
\be
\frac{\ddot a}{a}=\frac{\lambda}{3}-\frac{\omega \dot\phi^2}
{3\phi^2}-\frac{\dot a}{2a}\frac{\dot\phi}{\phi}
-\frac{\ddot\phi}{2\phi}-\frac{8\pi}{3\phi}\rho_{m}
+\frac{\Gamma}{2\phi}\frac{\dot\phi^2}{1-\phi}\, ,
\label{ab2}
\ee
and
\begin{eqnarray}
\ddot{\phi} + 3H \dot{\phi} + \frac{\dot{\phi}^2}{2 (1-\phi)}
 +\frac{8\pi(3p_m-\rho_m)}{2\omega+3}-\frac{\Gamma}{2\omega+3}
 \frac{\phi\dot\phi^2}{(1-\phi)^2}
 &-&\frac{2\Gamma}{2\omega+3}\frac{\phi\ddot\phi}{1-\phi} \nonumber \\
 &=&4\beta \phi (1-\phi^2) .
\label{fb1}
\end{eqnarray}
The energy density and the pressure
of the field $\phi$ \footnote{With the following expressions of
$\rho_\phi$ and $p_\phi$, the two equations (9) and (10) can be
expressed as $H^2=\frac{8\pi}{3\phi}(\rho_{m}+\rho_\phi)-\frac{k}{a^2}$
and $\frac{\ddot a}{a}=-\frac{4\pi}{3\phi}(\rho_{m}+\rho_\phi+
3p_m+3p_\phi)$ respectively.}
are given by
\be
\rho_{\phi}=\frac{1}{8\pi} \left(\phi \lambda
+ \frac{\omega \dot{\phi}^2}{2\phi}-3 H \dot{\phi}
-\frac{3}{2}\frac{\Gamma\dot{\phi}^2}{1-\phi}\right) ,
\label{fb2}
\ee
\be
p_{\phi} = \frac{1}{8\pi} \left(-\phi \lambda
+ \frac{\omega\dot{\phi}^2}{2\phi}+2 H \dot{\phi}+\ddot{\phi}
-\frac{1}{2}\frac{\Gamma\dot{\phi}^2}{1-\phi}\right) .
\label{fb3}
\ee
The energy equation can be found from the identity $(R^{\mu\nu}-
\frac{1}{2} g^{\mu\nu} R)_{;\nu}\equiv 0$, it is
\be
\dot{\rho}+3H(\rho+p)-\frac{\dot{\phi}}{\phi}\,\rho=0\, , \ \ \   \ \
\   \ \ \ (\rho=\rho_m+\rho_\phi ,\ \ \  \ \ \  p=p_m+p_\phi)\, .
\label{mb1} \ee In turn, the energy equation of matter is given by
\be
\dot{\rho}_m + 3 H(\rho_m + p_m)
=\frac{\Gamma}{4\pi}\,\left(\frac{\dot\phi\ddot\phi}{1-\phi}
+\frac{1}{2}\frac{\dot{\phi}^3}{(1-\phi)^2}+3H\frac{\dot\phi^2}{1-\phi}
\right) .
\label{mb2}
\ee
The terms in the right-hand side are responsible for the creation of
matter by gravitation characterized by the parameter $\Gamma$.
Heating is mainly in the later period of inflation  when $\dot\phi$
and $\ddot\phi$ get larger. When the $\phi$ oscillates after inflation,
$\dot\phi$ and $\ddot\phi$ have opposite phases, particle creation is
insignificant. The value of $\Gamma$ is determined by observations of
the present universe to be: $\Gamma/\xi=8.482\times10^{-3}$ (see section \ref{model}).
The temperature at the end of inflation is $\sim 10^{14}GeV$, there is no
need of reheating.

\sk

From equation (\ref{fb1}) we see that the scalar $\phi$ moves in
a double-well potential
\be
V(\phi) = \beta (1-\phi^2)^2
\label{fb4}
\ee
which ensures an asymptotic value $\phi_{a} = 1$ of $\phi$.
We note that this double-well is provided by the two functions
$\lambda(\phi)$ and $\omega(\phi)$ together \footnote
{It is given by $V'(\phi)=(2\phi^2\lambda'-2\phi\lambda)/(2\omega+3).$}
\vspace{0.5cm}.

\bsk

\section{THE VERY EARLY UNIVERSE: INFLATION}\label{infl}

\subsection{The creation of the universe}

The quantum creation of the universe starts at $t=0$ when the
gravitational scalar $\phi(t)=\phi(0)=0$. In fact, the probability
for creating the universe is given by $\cite{Linde}$
\be
P\sim e^{-3\phi^2/8V(\phi)} , \label{cu0}
\ee
where $V$ is the
scalar potential, so, this occurs to the highest local maximum
which locates at $\phi=0$ in the double-well potential. Or, we may
argue that $\cite{Pollock}$ the creation of the universe occurs at
the zero value of the action (zero action principle), and this is
the case $\phi=\partial\phi=0$ for the action (\ref{st1}).

We are also convinced that the universe is created by gravity from "nothing".
Before the creation of universe, there is no any matter but the field $\phi$.
If the field $\phi$ takes negative value and is trapped in the left
well of the potential, the equation $d^2 \phi/d\tau^2+4\beta(1-\phi^2)=0$
has the instanton solution
\be
\phi(\tau)=\tanh\,\sqrt{2\beta}\tau ,
\label{cu1}
\ee
where $\tau$ is the Euclidian time.

\sk

At $\phi=\dot\phi=0$, for $k=+1$, the Euclidean equation
\be
\left(\frac{da}{d\tau}\right)^2+\frac{2}{3}\xi\beta a^2-1=0
\ee
has the instanton solution
\be
a(\tau)=\frac{1}{\chi}\,\cos\,\chi \tau \; ,
\ee
here $\tau$ is the Euclidean time. This solution describes the creation
of a closed universe from  tunneling. At $\tau=0$, $a=1/\chi$.
By that time, equation (\ref{ab1}) has the classical solution
\be
a(t)=\frac{1}{\chi}\,\cosh\,\chi t\; ,
\label{cu2}
\ee
where
\be
\chi\equiv\sqrt{2\xi\beta/3}\; .
\label{cu3}
\ee

\subsection{The initial condition}

The universe after its creation experiences a quantum era
whence matter is created.  As this quantum era is known
very little, we shall set the initial conditions at the Planckian epoch
$t_0$ such that classical description is adequate hence forward.
At the Planckian epoch,
we can adopt the Planck relations: $l_{Pl}=\sqrt{G}$ and
$t_{Pl}=\sqrt{G}$,  and replace the gravitational
constant $G$ by $\phi^{-1}$. It is reasonable to let
$a_0\equiv a(t_0)\simeq \chi^{-1}$, then the initial condition follows as
\be
t_0=a_0\simeq\chi^{-1}\, ,  \ \ \  \phi_0\simeq \chi^2 .
\label{ini1}
\ee

\sk

The value of $\rho_{m0}$ is set to satisfy the causal constraint
$a_0H(t_0)\leq 1$. In general, $\rho_{m0}$ and $\rho_{\phi0}$
are of the same order. By the way, the exact value of initial
$\rho_m$ is of no importance, since it is damped away by inflation.

\sk

\subsection{Inflation}

In the very early universe, $\phi\sim 0$,
we see from eqs. (\ref{fb2},\ref{fb3}) that $p_\phi=-\rho_\phi$.
At the time $t_0$, $\rho_\phi$ and $\rho_m$ are of the same order,
there exists $3p+\rho<0$, therefore inflation commences already.
From this time on, $\rho_m$ decreases rapidly by redshift,
$\rho_\phi$ increases and soon dominates, there exists $p=-\rho$,
then the universe inflates exponentially with a Hubble parameter:
\be
H=\chi_e \; .
\label{inf1}
\ee
The constant $\chi_e$ is somewhat different from $\chi$ given in (18).
Since, in the slow-rolling approximation, the term $-H\dot\phi/\phi$
in equation (\ref{ab1}) is just $-4\beta(1-\phi^2)/3$, so that
the $2\xi\beta$ in $\lambda(\phi)$ should reduce to $(2\xi-4)\beta$,
we set it to be $(2\xi-3)\beta$ to
take account of the $\omega\dot\phi^{2}/6\phi^{2}$ term. Thus we take
\be
\chi_e=\sqrt{2\xi_e\beta/3}\; ,\ \ \  \ \ \  \xi_e=\xi-1.6\; .
\label{inf2}
\ee
The solutions of eqs. (\ref{ab1},\ref{fb1}) are
\be
a(t)=a_{0}\,e^{\chi_{e}(t-t_0)} ,
\label{inf3}
\ee
\be
\phi(t)=\phi_0e^{D(t-t_0)}
\label{inf4a}
\ee
respectively, where
\be
 D=(\sqrt{1+8/(3\xi_e)}-1)\sqrt{3\xi_e\beta/2}\; .
\label{inf4b}
\ee
In the above solutions, we have set $t_0=0.$
In equation (\ref{inf4a}), we have deleted the decaying solution.
Substituting the solution (\ref{inf4a}) into equation (\ref{mb2}) and
integrating, we have
\be
\rho_m=\frac{\Gamma r^2 (3H+r)}{4\pi (4H+2r)}\phi^2(t) ,
\label{inf5} \ee we have deleted the decaying solution also.
$\rho_m$ increases with $\phi^2$ steadily, there is no reheating
problem.

\sk

When $\phi(t)$ rolls down the potential hill,
all of $\phi$, $\dot\phi$ and $\ddot\phi$ increase, the energy
density of the field $\phi$ dominates continuously. When $\phi$
gets larger than $0.4$ or so,  $\rho_\phi+p_\phi$ becomes
positive and increases as $\phi$ increases. When $\phi$
gets to $0.9$ or so, it happens $\rho_\phi+3p_\phi=0$,
it is the same that $\rho+3p=0$, then inflation ends naturally.
The e-fold number of inflation is thus
\be
N=\frac{\chi_e}{D}(ln\,0.9-ln\,\phi_0) .
\label{inf6}
\ee

We may compare calculations from the above formulae with computer solutions
of the simultaneous equations (\ref{ab1}, \ref{fb1}, \ref{mb2}) given in
Table 1. Here we set

\begin{center}
$\xi=7.5\, ,\ \ \  \ \ \  \beta=1.7\times 10^{-16} , \ \ \  \ \ \
\Gamma=0.06362 . $
 \end{center}
The initial conditions according to the above subsection are
\begin{center}
$\phi=2.55\times 10^{-15} , \ \ \  \ \ \
\dot\phi=2.5\times10^{-23} ,\ \ \  \ \ \  \rho_m= 10^{-30} , \ \ \ \
\ \ a=2\times 10^7 . $
\end{center}
The calculated Hubble parameter, the ratio $\dot\phi/\phi$ and the
e-fold number are

\begin{center}
$\chi_e=2.5859\times 10^{-8} , \ \ \  \ \ \   D=7.9508\times
10^{-9} , \ \ \  \ \ \  N=109 .$
\end{center}

\sk

These values calculated from equations (\ref{inf2}\,--\,\ref{inf6}) can be
compared with the numerical solutions which are compiled
in Table I.

\sk

$\circ$ The calculated e-fold number is $109$ and that from numerical
solutions is $110$.

\sk

$\circ$ The Hubble parameter $H$ is nearly constant from Hubble
time $7$ to $90$ because of the pressure and the energy density of
field $\phi$ are equal in magnitude and opposite in sign and
$\rho_\phi > 10^5 \times \rho_m$.

\sk

$\circ$ The value D, i.e., $\dot\phi/\phi$, are nearly constant in the
period mentioned above, that is to say, $\phi$ increases exponentially
indeed.

\sk

$\circ$ The initial $\rho_m$ is damped away by $10$ Hubble times.
In the period $10-90$ Hubble time, the $\rho_{m(cal)}$ is less than
that of numerical solutions a little because of we have
ignored the second term in the right
bracket of equation ( \ref{mb2}) and $1-\phi$ is replaced by 1.
$\rho_m$ increases with $\phi^2$ in this period.

\sk

$\circ$ The temperature at the end of inflation is $3.5\times
10^{13} \,GeV$, this is the highest temperature of the universe.
Therefore GUT is never exact and there is no monopole problem at
all since monopoles are supposed to be produced when the exact GUT
is broken.

\bsk

%\pagebreak
\begin{center}
Table 1.  Numerical solutions of equations
(\ref{ab1},\ref{fb1},  \ref{mb2}).\\
\vspace{0.3cm}
\begin{tabular}{|l|c|c|c|c|c|}\hline
$t_{Hub}$   &$H$ &$\phi$  &$\dot\phi/\phi$ &$\rho_\phi$ &$\rho_{m}$
\\ \hline &&&&&\\
$1$  &$1.9429\cdot 10^{-8}$  &$4.9047\cdot 10^{-15}$
    &$1.1024\cdot 10^{-8}$ &$3.9893\cdot 10^{-31}$  &$1.7903\cdot 10^{-32}$\\
$4$  &$2.4903\cdot 10^{-8}$  &$7.4088\cdot 10^{-15}$
    &$8.4789\cdot 10^{-9}$ &$5.8882\cdot 10^{-31}$  &$3.3940\cdot 10^{-34}$\\
$7$  &$2.5848\cdot 10^{-8}$  &$3.5084\cdot 10^{-14}$
    &$7.9534\cdot 10^{-9}$ &$2.7981\cdot 10^{-30}$  &$6.9693\cdot 10^{-43}$\\
&&&&&\\
$10$  &$2.5848\cdot 10^{-8}$  &$8.8348\cdot 10^{-14}$
     &$7.9534\cdot 10^{-9}$ &$7.0461\cdot 10^{-30}$  &$1.7957\cdot 10^{-45}$\\
$20$  &$2.5848\cdot 10^{-8}$  &$1.9166\cdot 10^{-12}$
    &$7.9534\cdot 10^{-9}$  &$1.5285\cdot 10^{-28}$  &$8.4303\cdot 10^{-43}$ \\
$30$  &$2.5848\cdot 10^{-8}$  &$4.1549\cdot 10^{-11}$
    &$7.9534\cdot 10^{-9}$  &$3.3137\cdot 10^{-27}$  &$3.9622\cdot 10^{-40}$ \\
$40$  &$2.5848\cdot 10^{-8}$  &$9.0219\cdot 10^{-10}$
    &$7.9534\cdot 10^{-9}$  &$7.1953\cdot 10^{-26}$  &$1.8681\cdot 10^{-37}$\\
$50$  &$2.5848\cdot 10^{-8}$  &$1.9559\cdot 10^{-8}$
    &$7.9534\cdot 10^{-9}$  &$1.5599\cdot 10^{-24}$  &$8.7799\cdot 10^{-35}$\\
$60$  &$2.5849\cdot 10^{-8}$  &$4.2402\cdot 10^{-7}$
    &$7.9333\cdot 10^{-9}$  &$3.3817\cdot 10^{-23}$   &$4.1265\cdot 10^{-32}$\\
$70$  &$2.5850\cdot 10^{-8}$  &$9.207\cdot 10^{-6}$
    &$7.9530\cdot 10^{-9}$  &$7.3435\cdot 10^{-22}$  &$1.9453\cdot 10^{-29}$\\
$80$  &$2.5872\cdot 10^{-8}$  &$1.9945\cdot 10^{-4}$
    &$7.9477\cdot 10^{-9}$  &$1.5935\cdot 10^{-20}$  &$9.1192\cdot 10^{-27}$\\
$90$  &$2.6143\cdot 10^{-8}$  &$0.00422587$
    &$7.8820\cdot 10^{-9}$  &$3.4475\cdot 10^{-19}$  &$4.0464\cdot 10^{-24}$\\
$100$  &$2.7722\cdot 10^{-8}$  &$0.0753908$
    &$7.4583\cdot 10^{-9}$  &$6.9145\cdot 10^{-18}$  &$1.2467\cdot 10^{-21}$\\
&&&&&\\
 $105$  &$2.7483\cdot 10^{-8}$  &$0.76283$
    &$6.9723\cdot 10^{-9}$  &$2.4894\cdot 10^{-17}$  &$1.8678\cdot 10^{-20}$\\
$110$  &$1.1340\cdot 10^{-8}$  &$0.939783$
    &$2.3104\cdot 10^{-9}$  &$1.4136\cdot 10^{-17}$  &$2.9037\cdot 10^{-19}$\\
&&&&&\\ \hline
\end{tabular}
\end{center}

\section{THE UNIVERSE AFTER INFLATION}\label{after}

\subsection{Basic equations after inflation}

After inflation, let $\phi=1-\sigma^2$. We consider the evolution
for large time, $t\gg\beta^{-1}(\sim O(10^{-35})\,sec.)$,
$\sigma\ll1$. Equations (\ref{ab1}) and (\ref{fb1}) become \footnote{
The exact expression is
 $H^2=\frac{2}{3}\xi\beta (\sigma^2-(1-\sigma^2)ln(1-\sigma^2))+
    \frac{\xi\dot\sigma^2}{3(1-\sigma^2)}-\frac{\sigma^2\dot\sigma^2}
    {1-\sigma^2}+2\frac{\dot a}{a}\frac{\sigma\dot\sigma}{1-\sigma^2}+
    \frac{8\pi}{3}\frac{\rho_m}{1-\sigma^2}-\frac{2\Gamma\dot\sigma^2}
    {1-\sigma^2}-\frac{1}{a^2} .$  }
\begin{eqnarray}
H^2 &=& \frac{4}{3}\xi\beta\sigma^2+\left(\frac{\xi}{3}-2\Gamma\right)
\,\dot\sigma^2+\frac{8\pi}{3}\rho_m-\frac{1}{a^2} \nonumber \\
    &\approx & \frac{4}{3}\xi\beta\sigma^2+\left(\frac{\xi}{3}-
    2\Gamma\right)\,\dot\sigma^2+\frac{8\pi}{3}\rho_m ,
\label{aft1}
\end{eqnarray}

\begin{eqnarray}
\left(1-\frac{2\Gamma}{\xi}(1-\sigma^2)\right)\ddot\sigma+
    3H\dot\sigma+4\beta\sigma
  &=&6\beta\sigma^{3}-2\beta\sigma^{5}-\frac{4\pi}{3}
    \sigma(\rho_m-3p_m) \nonumber \\
  &\approx & 0 .
\label{aft2}
\end{eqnarray}
The energy density and pressure of $\phi$ are
\be
\rho_{\phi}=\frac{1}{8\pi}(4\xi\beta\sigma^2+\xi\dot\sigma^2
+6H\sigma\dot\sigma-6\Gamma\dot\sigma^2) ,
\label{aft3}
\ee
\be
p_{\phi}=\frac{1}{8\pi}\left(4\beta(2-\xi)\sigma^2+
(\xi-2)\dot\sigma^2
+2H\sigma\dot\sigma-\frac{2\Gamma(\xi-2)}{\xi}\dot\sigma^2\right) ,
\label{aft4}
\ee
and the equation (\ref{mb2}) becomes
\be
\dot\rho_{m}+3H(\rho_{m}+p_{m})=\frac{\Gamma}{\pi}
\,(\dot\sigma\ddot\sigma+3H\dot\sigma^2) .
\label{aft5}
\ee

Since the energy density $\rho_{\phi}$ dominates and it is nonrelativistic
\footnote{When $\phi(t)$ gets near the point 1, the bottom of the
potential well, $\phi$ behaves as particles with effective mass $\sim$
2$\sqrt\beta$ larger than its kinetic energy.} ,
the universe expands by power law about $2/3$. From equation (\ref{aft2}),
we see that $\sigma$ oscillates as
\be
\sigma(t)=f(t)\,\cos(\varpi t) , \ \ \  \ \ \  \ \ \
\varpi^2=4\beta/(1-2\gamma) ,
\label{sg1}
\ee
where
\be
\gamma=\frac{\Gamma}{\xi}\; .
\label{sg2}
\ee
By equation (\ref{aft2}), $f(t)$ satisfies
\be
\dot f=-\frac{3H}{2(1-2\gamma)}f\; .
\label{sg3}
\ee

\subsection{The radiation era: $p_m=\frac{1}{3}\rho_m$}
By numerical solutions on computer, we see that $f\sim \frac{1}{t}$.
Let
\be
f(t)=\frac{c}{t}\, ,
\label{re1}
\ee
and substituting into (\ref{sg3}),we get
\be
H=\frac{2(1-2\gamma)}{3 t}\, .
\label{re2}
\ee
Then, substituting $f(t)$ and $H(t)$ into equation (\ref{aft5})
and integrating, we get solutions:
\be
\sigma(t)=\sqrt{\frac{1-8\gamma}{1-4\gamma}}\,
\frac{1-2\gamma}{\sqrt{3\xi\beta}}\,\frac{\cos(\varpi t)}{t}\, .
\label{re3}
\ee
The densities are
\be
\rho_m=\frac{(1-2\gamma)\gamma}{\pi t^2}\, ,
\label{re4}
\ee
and
\be
\rho_\phi=\frac{(1-2\gamma)(1-8\gamma)}{6\pi t^2}\, .
\label{re5}
\ee
In the above solutions, the constant $c$ has been determined by
equation (\ref{aft1}).

\subsection{The dust era: $p_m=0$}

After the time $t_{eq}$, the time of radiation-dust equality,
$p_m=0$, the numerical solutions show that
\be
H=\frac{2}{3t}\; .
\label{de1}
\ee
From (\ref{sg3}), we have
\be
f(t)\sim \frac{1}{t^{1/(1-2\gamma)}}\; .
\label{de2}
\ee
After integrating (\ref{aft5}) and determining the integrating constant,
we have solutions:
\be
\sigma(t)=\sqrt{\frac{1-8\gamma}
{1-4\gamma}}\frac{1}{\sqrt{3\xi\beta}}\frac{t_{eq}^{2\gamma/
(1-2\gamma)}}{t^{1/(1-2\gamma)}}\,\cos(\varpi t)\; ;
\label{de3}
\ee
\be
\rho_m=\frac{1}{6\pi t^2}\left(1-\frac{1-8\gamma}{1-2\gamma}
\left(\frac{t}{t_{eq}}\right)^{-4\gamma/(1-2\gamma)}\right) ,
\label{de4}
\ee
and
\be
\rho_\phi=\frac{1}{6\pi t^2}\,\frac{1-8\gamma}{1-2\gamma}
\left(\frac{t}{t_{eq}}\right)^{-4\gamma/(1-2\gamma)} .
\label{de5}
\ee
The function $\omega$ increases as $\sim t^2$. For the
given model, for example, at $t=1sec$, $\omega\sim O(10^{73})$.
The scalar-tensor gravity agrees with the general relativity
exactly.

\bsk

\section{THE MODEL}\label{model}
Now we determine the three parameters $(\xi,\beta,\Gamma)$ in the model
from observation data. The parameter $\Gamma$ can be determined by the
values of $h,\, T$ and $\Omega_m$ at present time $t_0$ as follows.

\subsection{Determining $\Gamma/\xi$}

We assume $\Omega_0=1$ and set $h=0.6$, then $t_0=6.361\times
10^{60}=10.87\,Gy$
and $\rho_c=1.312\times 10^{-123}=6.765\times 10^{-30}\,g/cm^3$.
The temperature is
determined to be $T_0=1.925\times 10^{-32}=2.728\,K$ and $\rho_\gamma=9.039
\times 10^{-128}=4.662\times 10^{-34}\,g/cm^3$.
The density of matter  is assumed to be
$\Omega_m=0.4$ and so $\rho_m=5.248\times 10^{-124}=2.706\times
10^{-30 }\,g/cm^3$. Thus the temperature of dust-radiation equality
is calculated to be
\be
T_{eq}=6.651\times 10^{-29}=9423\,K .
\label{ga1}
\ee
We have the first equation to determine the parameter $\Gamma$
and the time $t_{eq}$:
\be
\gamma (1-2\gamma)\,\left(\frac{t_0}{t_{eq}}\right)^2=2.747\times 10^9 ,
\label{ga2}
\ee
where $\gamma\equiv\Gamma/\xi.$ On the other hand,
relating the expression (\ref{de4}) for $\rho_m(t_0)$ to the
value $\rho_m=1.822\times 10^{-124}$ gives the second equation:
\be
1-\frac{1-8\gamma}{1-2\gamma}\left(\frac{t_0}{t_{eq}}\right)^{
-4\gamma/(1-2\gamma)}=0.4\; .
\label{ga3}
\ee
Solving the above two equations simultaneously gives
\be
\begin{array}{l}
t_0/t_{eq}=573900 , \ \ \  \ \ \  t_{eq}=1.108\times 10^{55}=18940\,yr , \\
\gamma=0.008482 .
\end{array}
\label{ga4}
\ee

\sk

It should be noted that this result depends only on the adapted values of
$h,T_0, \Omega_m$ but not others.

\subsection{The spectra of density perturbations and gravitational waves}

Since the energy density of the field $\phi$ dominates in
the evolution of the universe, we consider its adiabatic density
perturbations. We perform a conformal transformation:
\be
g_{\mu\nu}=A^2(\tilde\phi)\tilde g_{\mu\nu} ,\ \ \  \ \ \
 A^2(\tilde\phi)=\frac{1}{\phi}\; ,
\label{spectra1}
\ee
and
\be
\tilde\phi=\sqrt{\frac{\xi}{16\pi}}\,ln\frac{1-\sqrt{1-\phi}}
{1+\sqrt{1-\phi}} ,
  \ \ \   \ \ \  \ \ \  \frac{d\tilde\phi}{d\phi}=\sqrt{\frac{\xi}
  {16\pi}}\frac{1-\phi}{\phi} ,
\label{spectra2}
\ee
then
\be
{\cal A}=\int d^4x\sqrt{-\tilde{g}}\,[\frac{1}{16\pi}R(\tilde{g})-\frac{1}{2}
  \tilde{g}_{\mu\nu}{\tilde\phi}^{,\mu}
   {\tilde\phi}^{,\nu}-2\tilde{V}(\tilde\phi)] ,
\label{spectra3}
\ee
where
\be
\tilde{V}(\tilde\phi)=\frac{\lambda(\phi)}{8\pi\phi}\; .
\label{spectra4}
\ee

\sk The field $\tilde\phi$ is a canonical field, then we can
calculate the slow-roll parameters
\be
\epsilon\equiv\frac{1}{16\pi}\left(\frac{V'(\tilde\phi)}{V(\tilde\phi)}
\right)^2=\frac{1}{\xi}\; ,
\label{spectra5}
\ee
and
\be
\eta\equiv\frac{1}{8\pi}\frac{V''(\tilde\phi)}{V(\tilde\phi)}
=\frac{1}{\xi}\; .
\label{spectra6}
\ee

The spectrum of adiabatic density perturbation is
\be
\delta^2_H(k)=\frac{512\pi}{75}\frac{V^3(\tilde\phi)}{V'^2(\tilde\phi)}
=0.03395\,\frac{\beta\xi^2}{\phi} .
\label{spectra7}
\ee
In these results, we have noticed that $\phi\ll 1$ for interested
$k$ in the range $1/h^{-1}Mpc\,-\, 1/3000h^{-1}Mpc$.

The spectral index is
\be
n_s=1+\frac{d\,ln\,\delta^2_H(k)}{d\,ln\,k}=1-\frac{D}{H}=1-p\; .
\label{spectra8}
\ee
where $p\equiv D/H=\frac{3}{2}\left[(1+8/3\xi_e)^{1/2}-1\right]$ which
depends on $\xi$ only.
We note that $d\,ln\,k\sim Hdt$, since
at horizon crossing $k=aH$ and $a\sim e^{Ht}$, $H$
is constant.

\bsk Inflation also generates gravitational waves, whose relative
contribution to the mean-squared low multipoles of the CMB
anisotropy is
\be
r_{ts}\equiv
\frac{5}{8\pi}\left(\frac{V'(\tilde\phi)}{V(\tilde\phi)}\right)^2
=10\epsilon\; . \label{spectra9}
\ee
This more correct factor $10$
was given by Polarski and Starobinsky $\cite{Polarski}$.
For our model, $r_{ts}=1.3$, this rather large value should be
tested by future observations.

\subsection{COBE}

According to COBE DMR observations of large-angular-scale
CMB anisotropy$\cite{Bunn}$, we normalize the density perturbation spectrum
using the result$\cite{White}$:
\be
\delta_H(H_0)= 1.91\times 10^{-5}\,\frac{\exp{[1.01\,(1-n_s)]}}
{\sqrt{1+0.75r_{ts}}} .
\label{co1} \ee
Let $n_s=0.7\, \footnote{Smaller $n_s$ gives smaller $(\xi,\beta)$ but larger
large-scale streaming velocities. For example, $n_s=0.55,$  gives \, $(\xi,\beta)$
=$(5.5,\,1.7\times 10^{-18})$,\,\,$\sigma_8=0.68,\,
\sigma_v(40)=451\,km/s,\,\sigma_v(60)=414\,km/s$.}
 \,\, i.e.,\,\,p=0.3 ,$ then
\be
\xi=7.5\; ,
\ee
and eqs.(\ref{spectra7}, \ref{spectra9}, \ref{co1})
determine
\be
\beta=1.7\times 10^{-16} .
\ee
In this model, the energy scale of inflation is
\be
\tilde V^{1/4}=(6.3\times 10^{16}\,GeV)\,\epsilon^{1/4} .
\ee

\sk To convince the viability of this model, we calculate
$\sigma_8$ and the large-scale streaming velocities $\sigma_v(r)$
which are the three dimensional velocity dispersions of galaxies
within sphere of radius $rh^{-1}Mpc$ after the data have been
smoothed with a Gaussian filter of $12\,h^{-1}Mpc$. The results
are given \footnote{The observational data $\cite{Bertschinger}$
are \hspace{5mm}
$\sigma_v(40)=388(1\pm 0.17)\,km/s$ \hspace{2mm} and \hspace{2mm}
$\sigma_v(60)=327(1\pm 0.25)\,km/s$.} in Table 2.

\vspace{0.3cm}
\begin{center}
Table 2.  Some parameters of the model

\vspace{0.3cm}
\begin{tabular}{|c|c|c|c|}\hline
  $n_s$ &$\sigma_8$
       &$\sigma_v(40)$ &$\sigma_v(60)$  \\  \hline
   $\hs 0.7\hs$ &$\hs 0.69\hs$ &$\hs 357\,km/s\hs$
   &$\hs 309\,km/s\hs$\\
\hline
\end{tabular}
\end{center}
\sk

We note that in calculating the streaming velocities,
we have adopted the CDM transition function \cite{Sagiyama}
\be
T(q)=\frac{ln\,(1+2.34 q)}{2.34 q}\left[1+3.89 q+(16.1 q)^2+
 (5.46 q)^3+(6.71)^4\right]^{-\frac{1}{4}}
\ee
with $q=k/h\Gamma_s\,(Mpc^{-1})$ , and $\Gamma_s=\Omega_0 h\ ,
exp(-\Omega_B-\sqrt{\frac{h}{0.5}}\Omega_B/\Omega_0)$ .
This manifests itself that
the gravitational scalar field $\phi$ is just the cold dark matter
men seek for, of course, there may be other dark matter contained
in $\Omega_m$ (in this model, $\Omega_m=0.4 ,\, \Omega_B h^2=0.019$) .

\section{THE PRIMORDIAL NUCLEOSYNTHESIS}\label{nucleo}

Now, we consider the constraint from primordial nucleosynthesis on
scalar-tensor theories of gravity. Nucleosynthesis is sensitive to the
speed-up factor $\xi_n $ which cannot differ from unity by more than
$\sim 15 \%$ without over- or under producing $^4$He $\cite{Santiago}$.
The $\xi_n$ has been derived to be
\be
 \xi_n=\frac{A(\varphi_n)}{\sqrt{1+\alpha^2(\varphi_0)}}
\ee
with $\varphi_n$ being the value of the scalar field at the time of
nucleosynthesis and $\varphi_0$ its value today. $A(\varphi)=1/\sqrt{\phi},$
$\alpha(\varphi)$ is defined by $[12]$
\begin{equation}
a(\varphi)\equiv ln\,A(\varphi) ,
\end{equation}
\begin{equation}
\alpha(\varphi)\equiv \frac{\partial a(\varphi)}{\partial(\varphi)} .
\end{equation}
Here, $\varphi=\sqrt{8\pi}\tilde\phi$, since $[12]$
\begin{equation}
{\cal A}=\int d^4x\sqrt{-\tilde{g}}\,\frac{1}{16\pi}\left[R(\tilde{g})
-\frac{1}{2}\tilde{g}_{\mu\nu}{\varphi}^{,\mu}
   {\varphi}^{,\nu}-2U(\varphi)\right] ,\ \ \  \ \ \
U(\varphi)=\frac{\lambda(\phi)}{\phi} .
\end{equation}
Therefore
\begin{equation}
\alpha(\varphi)=-\sqrt{\frac{1}{2\xi}}\sqrt{1-\phi} .
\end{equation}
At the time of nucleosynthesis, $A(\varphi_n)=1+O(10^{-72})$, and
at $t_0$, $\alpha^2(\varphi_0)$ is of order $O(10^{-104})$, then
\begin{equation}
\xi_n=1+O(10^{-72}) .
\end{equation}
We see that it makes no significant differences with the standard cosmology.

\end{document}